\documentclass[aps,prl,a4paper,twocolumn]{revtex4}

\usepackage{graphics}
\usepackage{amssymb}
\usepackage{amsmath}
\usepackage{hyperref}
\usepackage{xcolor}
\usepackage{physics}
\usepackage{graphicx}

\def\beq{\begin{equation}}
\def\eeq{\end{equation}}
\def\beqa{\begin{eqnarray}}
\def\eeqa{\end{eqnarray}}

\def\cB{{\cal B}}

\def\cH{{\cal H}}

\def\vr{{\mathbf{r}}}
\def\vx{{\mathbf{x}}}

\def\vp{{\mathbf{p}}}
\def\vq{{\mathbf{q}}}
\def\vk{{\mathbf{k}}}
\def\vs{{\mathbf{s}}}

\def\vtp{{\bf \tilde{p}}}
\def\vtq{{\bf \tilde{q}}}

\begin{document}


\title{Entanglement Renormalization for Interacting  
 Field  Theories} 

\author{J.J. Fern\'andez-Melgarejo$^{(1)}$}
\email{jj.fernandezmelgarejo@um.es}

\author{ J. Molina-Vilaplana$^{(2)}$}
\email{javi.molina@upct.es}

\author{ E. Torrente-Lujan$^{(1)}$}
\email{torrente@cern.ch} 

\affiliation{${}^{(1)}$ Dept. of Physics, Universidad de Murcia. Murcia, Spain} 
\affiliation{${}^{(2)}$ Universidad Polit\'ecnica de Cartagena. 
Cartagena, Spain.}

\date{\today}

\begin{abstract}
A general method to build the entanglement renormalization (cMERA) for interacting quantum field theories is presented.  We improve upon the well-known Gaussian formalism used in free theories through a class of variational non-Gaussian wavefunctionals for which expectation values of local operators can be efficiently calculated analytically and in a closed form. The method consists of a series of scale-dependent nonlinear canonical transformations on the fields of the theory under consideration.  {Here, the $\lambda\, \phi^4$ and the sine-Gordon scalar theories are used to illustrate how non-perturbative effects far beyond the Gaussian approximation are obtained by considering the energy functional and the correlation functions of the theory.}
\end{abstract}

\maketitle

In recent years, tensor networks, a new and powerful class of variational states, have proved to be very useful in addressing both static and dynamical aspects of a wide number of interacting many-body systems. They represent a class of systematic variational \emph{ans\"atze} which, through the Rayleigh-Ritz variational principle, provide an elegant approximation to the ground state of an interacting theory by systematically identifying those degrees of freedom that are actually relevant for observable physics. These variational \emph{ans\"atze} are nonperturbative  and can be applied both in the lattice and in the continuum. As an example, the Multiscale Entanglement Renormalization Ansatz (MERA), a variational real-space renormalization scheme on the quantum state, represents the wavefunction of the quantum system  at different length scales \cite{Vidal1}. 

A continuous version of MERA, known as cMERA, was proposed in \cite{cMERA1} for free field theories. It consists of building a scale-dependent representation of the ground state wavefunctional through  a scale-dependent linear canonical transformation of  the fields of the theory. Namely, the renormalization in scale is generated by a quadratic operator, and thus, the resulting state is given by a Gaussian wavefunctional. Despite this fact obviously limits the interest of this trial state for interacting {quantum field theories (QFT)}, the Gaussian ansatz has been used in cMERA and correctly reproduces correlation functions and entanglement entropy in free field theories \cite{cMERA2, Vidal2}. 
Furthermore, as the Gaussian cMERA is currently studied as a possible realization of holography \cite{HolCMERA1,HolCMERA2,HolCMERA3,HolCMERA4, JMV1, JMV2}, it is timely to develop interacting versions of cMERA in order to advance in this program. In \cite{Cotler1}, the Gaussian cMERA was applied to interacting bosonic and fermionic field theories. In \cite{Cotler2}, authors developed some techniques to build systematic perturbative calculations of cMERA circuits but restricted to the weakly interacting regime. 

Our aim here is to provide a non-perturbative method to build truly non-Gaussian cMERA wavefunctionals for interacting QFTs. A justifiable way of doing so would be to formulate a perturbative expansion for which the Gaussian wavefunction appears in its first order \cite{cluster2,cluster3,cluster4,cluster5}.  Unfortunately, with these methods, expectation values of operators cannot be calculated exactly and must be approximated by an additional series expansion.  On the contrary, our approach clings to the variational method, but using a more elaborated class of trial wavefunctionals. 
Here, we use a set of nonlinear canonical transformations (NLCT) \cite{polley89, ritschel90, ritschel91,ritschel94,ibanez} to build a set of scale-dependent extensive functionals which are certainly non-Gaussian. Remarkably, with this prescription, observables can be analytically calculated in a closed form.
{We illustrate the method by considering the self-interacting $\lambda \phi^4$ scalar theory and the sine-Gordon model 
in $(d+1)$ dimensions. For $d=1$, these theories, do not exhibit any issue when renormalization is considered, and thus the non-Gaussian cMERA lies on a solid ground.} In addition, our variational procedure adds up a much larger class of Feynman diagrams than the usual ``cactus''-like ones which are captured by the Gaussian approach \cite{stevenson}. Therefore, we are certainly generalizing the variational approach in QFT to non-Gaussian trial states in the canonical formalism. 
\\
%

\emph{Gaussian cMERA.-}
cMERA \cite{cMERA1, cMERA2} is a real-space renormalization group procedure on the quantum state that builds a scale-dependent wavefunctional $\Psi[\phi,u]$,
\begin{equation}
\label{cMERAansatz}
\Psi[\phi,u]=\langle \phi|\Psi_u\rangle = \langle \phi|\, \mathcal{P} \, e^{-i \int_{u_{\text{IR}}}^u (K(u') + L) \, du' } \, |\Omega_{\text{IR}}\rangle
\, ,
\end{equation}
where $u$ parametrizes the scale of the renormalization.  (\ref{cMERAansatz}) contains the path-ordered exponential of the dilatation operator $L$ and the generating operator $K(u')$.  The renormalization scale parameter $u$ in cMERA  is usually taken to be in the interval $[u_{\scriptscriptstyle IR},u_{\scriptscriptstyle UV}] = (-\infty,0]$.  $u_{\scriptscriptstyle UV} = u_{\epsilon}$ is the scale at the UV cutoff $\epsilon$, and the corresponding momentum space UV cutoff is $\Lambda = 1/\epsilon$.  $u_{\scriptscriptstyle IR} = u_{\xi}$ is the scale in the IR limit, where $\xi$ is a long-wavelength correlation length.  The state $|\Psi_{\scriptscriptstyle UV} \rangle$ is the ground state of a quantum field theory.   The $L$-invariant state $| \Omega_{\text{IR}} \rangle$ is a Gaussian state with no entanglement between spatial regions. The cMERA Hamiltonian evolution generates translations along the cMERA parameter $u$.  The term $K(u)$ in the cMERA-Hamiltonian is called the \emph{entangler} operator and the only variational parameters of the ansatz are those which parametrize it.  For free scalar theories, $K(u)$ is the quadratic operator given by \cite{cMERA1,cMERA2}
\begin{equation}
K(u) =
\frac{1}{2}\int_{\vp}\, g_0(p,u)\, \left[
	\phi(\vp)\pi(-\vp)
	+ \pi(\vp)\phi(-\vp)
	\right]
\, ,
\label{eq:entangler}	
\end{equation}
{where $p\equiv|\vp|$ and $\int_{\mathbf{p}}\equiv\int\, (2\pi)^{-d}\, d^{d}p$ with $d$, the spatial dimensions of the theory.  The conjugate momentum of the field $\phi(\vp)$ is $\pi(\mathbf{p})=-i\bar{\delta}/\delta\phi(-\vp)$, such that $[\phi(\mathbf{p}), \pi(\mathbf{q})] = i\bar{\delta}(\mathbf{p+q})\,$, with $\bar{\delta}(\mathbf{p})\equiv(2\pi)^{d}\delta(\mathbf{p})$. The function  $g_0(p,u)$ in (\ref{eq:entangler}) is the only variational parameter to optimize in the Gaussian cMERA.} This function factorizes as $g_0(p;u)=g_0(u)\, \Gamma(p/\Lambda)$ where  $\Gamma(x)\equiv \Theta(1-|x|)$ and $\Theta(x)$ is the Heaviside step function; $g_0(u)$ is a real-valued function and $\Gamma(p/\Lambda)$ implements a high-frequency cut-off such that $\int_{\vp} \equiv \int_{\vp}^{\Lambda}$. Choosing $|\Omega_{IR}\rangle$ as \cite{cMERA2}
\begin{align}
\label{scaleinv1}
\left(\sqrt{\omega_{\Lambda}} \,(\phi(\vp) - \chi_0 ) + \frac{i}{\sqrt{\omega_{\Lambda}}} \, \pi(\vp)\right) |\Omega_{IR}\rangle = 0
\, ,
\end{align}
$\text{for all }\vp\, ,$ where $\omega_{\Lambda}=\sqrt{\Lambda^2 + m^2}$ {with $m$ the mass of the particles in the free theory}, it is possible to show that the cMERA ansatz with a quadratic entangler is equivalent to the Gaussian wavefunctional
 given by 
\begin{align}
\label{cMERAwavefunctional1}
\Psi[\phi;u]_{SG} = 
N\, e^{-\frac{1}{2}\int_{\vp} \, \left(\phi(\vp) - \chi_0 \right) \, F^{-1}(p ; u) \, \left(\phi(-\vp) - \chi_0\right)} 
\, ,
\end{align}
 where {$\chi_0=\langle \Psi_{SG}(u=0) |\phi(x)|\Psi_{SG}(u=0)\rangle $} and the relation between the scale-dependent Gaussian kernel $F(p; u)$ and the variational cMERA parameter $g_0(p,u)$ is given by \cite{Cotler1}
\begin{align}
\label{dictionaryJMV}
F^{-1}(p;u) = \omega_\Lambda \, e^{2 \int_0^u d u' \ g_0(p e^{-u'}, u')} \,,
\end{align}
with {$F(p;0)=(p^2 + m^2)^{-1/2}$.} 


We note that $\Psi[\phi;u]_{SG} = U_S\Psi_G[\phi;u]$, where the operator that shifts the argument of any functional (and specifically the Gaussian wavefunctional) by a constant $\chi_0$, is given by $U_S = e^{O_S}$ with $O_S=-\int_{\vp}\chi_0\, \delta/\delta\phi(-\vp)$. Then, defining $U_G(u_1,u_2)\equiv \mathcal{P}e^{ -i\int_{u_2}^{u_1} du (K(u)+L)}$, the Gaussian state is given by $\Psi[\phi;u]_{G}= \bra{\phi} U_G(u,u_{IR})\, |\Omega_{IR}\rangle$.

Finally,  we remark that the Gaussian cMERA ansatz may be also understood as the set of scale-dependent linear transformation of the fields given by 
\begin{align}
U_G(0,u)^{-1} \phi(\vp)U_G(0,u)
=&\
e^{-f(p,u)}e^{-\frac{u}{2}d}\phi(\vp e^{-u})
\ ,
\\
U_G(0,u)^{-1} \pi(\vp)U_G(0,u)
=&\
e^{f(p,u)}e^{- \frac{u}{2}d}\pi(\vp e^{-u})\, ,
\end{align}
with $f(p,u) = \int_0^u d u' \ g_0(p e^{- u'}, u')$.
\\

\emph{Non-Gaussian cMERA.-}
In QFT, trial states created by introducing polynomial corrections to a Gaussian state correspond to a finite number of particles and those are suppressed in the thermodynamic limit. Thus, in going beyond the Gaussian ansatz, it is necessary to use a class of variational extensive states for which the energy density does not depend on the volume. Following \cite{polley89,ritschel90,ibanez}, we build extensive non-Gaussian trial states considering wavefunctionals of the form
\beq
{\Psi_{NG}[\phi] =U_{NG}\, \Psi_G[\phi]=\exp(\cB)\, \Psi_G[\phi] \, ,}
\eeq
where the $NG$ subscript refers to non-Gaussian, $\Psi_G[\phi]$ is a normalized Gaussian wavefunctional and $U_{NG}=\exp(\cB)$, with $\cB^{\dagger} = -\cB$, an anti-Hermitian operator that, for the moment, it may add new variational parameters, in addition to those in the Gaussian wavefunctional. The expectation value of any operator $\mathcal{O}(\phi,\pi)$ in these states amounts to the calculation of a Gaussian expectation value for the transformed operator $\widetilde{\mathcal{O}}=U_{NG}^{\dagger}\, \mathcal{O}\, U_{NG}$, i.e., {$\langle \Psi_{NG} |\mathcal{O}(\phi,\pi) | \Psi_{NG}\rangle = \langle \Psi_{G} |U_{NG}^{\dagger}\, \mathcal{O}(\phi,\pi)\, U_{NG} | \Psi_{G}\rangle\, .$}
The transformed operator $\widetilde{\mathcal{O}}$ is straightforwardly built once the transformations 
\beq
\widetilde{\phi}(\vp) = U_{NG}^{\dagger}\, \phi(\vp)\, U_{NG}\, , \quad \widetilde{\pi}(\vp) = U_{NG}^{\dagger}\, \pi(\vp) U_{NG}\, ,
\eeq
are known. 
The  transformation on the  operator $\mathcal{O}$ generated by $\cB$ is given by the Hadamard's lemma in 
terms of a series of nested commutators
\footnote{${\rm Ad}_{\cB}\, (A)\equiv \exp(\cB)\, A\, \exp(-\cB)$,  ${\rm ad}_{\cB}(A)\equiv [\cB,A]$.} 
 \beq
\widetilde{\mathcal{O}}= {\rm Ad}_{\cB}\, (\mathcal{O})= e^{{\rm ad}_{\cB}}\, \mathcal{O}\, .
\label{eq:conm_series}
\eeq
It can be seen that  a suitable choice of $\cB$, while  leading  to a non-Gaussian trial state, can indeed truncate the commutator expansion, thus reducing the calculation of expectation values of functionals to a finite number of Gaussian expectation values 
\footnote{In \cite{Cotler3}, an analysis of the Bender-Dunne basis has been done and some non-linear transformations are proposed. Because nested commutators with $n\ge2$ induce operators of powers higher than 4, the algebra closes perturbatively. Hence, their construction is done up to ${\cal O}(\lambda)$, namely for weak coupling regimes.}. 
The exponential form of the transformation ensures the correct extensive volume dependence of observables such as the energy of the system. In addition, as $U_{NG}$ is unitary, the normalization of the state is preserved. The operator $\cB$ consists of a product of $\pi$'s and $\phi$'s, which is given by
\beqa
\cB 
&=& 
-s\int_{\vp\, \vq_1\cdots \vq_m} \hspace{-0.8cm}
		h(\vp,\vq_1,\ldots,\vq_m)\frac{\delta}{\delta \phi(-\vp)}\phi(\vq_1)\ldots\phi(\vq_m) \nonumber
\ ,
\eeqa
where $h(\vp,\vq_1,\ldots,\vq_m)=g(p,q_1,...,q_m) \bar{\delta}(\vp+\vq_1+\cdots+\vq_m) $, $s$ is a variational parameter, $g(p,q_1,\ldots,q_m)$ is a variational function that must be optimized upon energy minimization and $m\in \mathbb{N}$. 
The other variational parameter is the kernel $F(p)$ entering the Gaussian wavefunctional. {The explicit dependence of these  parameters on the interaction couplings of a theory is established through energy minimization. This will be discussed later for some concrete examples.} The function $g(p,q_1,\ldots,q_m)$ is symmetric under the exchange of $q_i$'s,  it must ensure the anti-Hermiticity of $\mathcal{B}$ and is constrained to satisfy
$g(p,p,q_2,\ldots,q_m)=0$ and $g(p,q_1,\ldots,q_m)g(q_i,k_1,\ldots,k_m) 
= 0$, for $i=1,\ldots,m$.
This constraint ensures that the multiple commutator series in (\ref{eq:conm_series}) terminates after the first non-trivial term. Such procedure yields a variational approximation to the calculation of observables in an interacting theory which improves upon the Gaussian ansatz. The parameter $s$ is a truly non-Gaussian tracking parameter which shows the deviation of any observable from the Gaussian case. 
 
The action of $U_{NG}$ on the canonical field operators is given by
\begin{align}
\widetilde{\phi}(\vp)=
\phi(\vp)
+s\,  \Phi(\vp) 
\ ,
\quad
\widetilde{\pi}(\vp)=
\pi(\vp)
 + s\, \Pi(\vp) 
\ ,
\label{eq:nlct}
\end{align}
with
\beqa
\Phi(\vp)
\hspace{-0.1cm}&=&\hspace{-0.1cm}
 \int_{\vq_1\cdots \vq_m} \hspace{-0.8cm}
		h(\vp\, , -\vq_1\cdots -\vq_m)\phi(\vq_1)\cdots \phi(\vq_m) 
\ ,   \\
\Pi(\vp) \hspace{-0.1cm}&=&\hspace{-0.1cm} -m \int_{\vq_1\cdots \vq_m} \hspace{-0.8cm}
		h(-\vq_1\, ,\vp, \cdots -\vq_m)\pi(\vq_1)\phi(\vq_2)\cdots \phi(\vq_m) \nonumber
\ .
\eeqa
The canonical commutation relations (CCR)  still hold under the unitary, albeit non-linear,
 transformation of the fields \eqref{eq:nlct},  $[\widetilde{\phi}(\vp), \widetilde{\pi}(\vq)]=i\bar{\delta}(\vp + \vq)\, .$
Noticing that the Gaussian cMERA is generated by the quadratic operator (\ref{eq:entangler}), it is clear 
that operators $\cB$ which are linear or quadratic in $\pi$'s and $\phi$'s do not yield any improvement upon the Gaussian ansatz. Therefore, in going beyond,  one must consider operators $\cB$ that at least are cubic in the products of these fields.

{In terms of wavefunctionals, 
the action of $U_{NG}$ on a functional  $A[\phi]$ can be understood as a nonlinear shifting of the argument from $\phi$  to $\phi-s\, \phi^{m}$ and thus, for the Gaussian wavefunctional, $U_{NG}\, \Psi_{G}[\phi]= \Psi_{G}[\phi-s\, \phi^{m}]\, $ \cite{polley89, ritschel90}.}

Hence, our proposal to build non-perturbative cMERA states for interacting field theories is based on the idea of defining the set of scale-dependent non-linear transformations 
\beqa
\widetilde{\phi}(\vtp) &=&U_{NG}(u)^\dagger \phi(\vp) U_{NG}(u) \ ,\\
\widetilde{\pi}(\vtp)&=&
U_{NG}(u)^\dagger \pi(\vp) U_{NG}(u)
\ ,
\eeqa
where $U_{NG}(u) \equiv U_{NG}\, U_{SG}(u)\,$ and $ U_{SG}(u)\equiv U_S\,  U_G(u,u_{IR})$. As commented above, in going beyond the Gaussian approach,  for $U_{NG}$ one must consider operators $\cB$ that at least are cubic in the products of these fields. Here we will focus in the simplest one \footnote{{Namely, the NLCT method allows to build a set of non-Gaussian cMERA tensor networks, each one based on different choices of $\cB = \pi\, \phi^m$, with $m \in \mathbb{N}$.} }, i.e., the case $m=2$ which {we denote by $\cB = \pi\, \phi^2$} and explicitly reads 
\begin{align}
\cB 
= 
-s\int_{\vp\, \vq_1, \vq_2} \hspace{-0.6cm}
		g(p,q_1,q_2)\pi(\vp)\phi(\vq_1)\phi(\vq_2) \bar{\delta}(\vp+\vq_1+\vq_2) 
 ,
\end{align}
where, from a cMERA point of view, $g(p,q_1,q_2)$ can be interpreted as a variational coupling-dependent momentum cut-off function \cite{supp_material}. With this choice for $\cB$, the transformed fields result
\beqa
\label{eq:nlct_u}
\widetilde{\phi}(\vtp)&=&
\Sigma_{(-)}(\tilde p e^{u};u)\left(
	\phi(\vtp)
	+s e^{\frac d2 u}\, \Phi(\vtp)
\right)
\ , \\ \nonumber
\widetilde{\pi}(\vtp)&=&
\Sigma_{(+)}(\tilde p e^{u};u)\left(
	\pi(\vtp)
	-2s e^{\frac d2 u}\, \Pi(\vtp)
\right)
\ ,
\eeqa
where we have made the change of variables in momenta $\vp \equiv e^u \vtp\,$. In addition, we have defined $\Sigma_{(\pm)}(p;u) \equiv e^{\pm f(p,u)}e^{-\frac d2 u}$ and
\begin{eqnarray}
\Phi(\vtp)&=&
\int_{\vtq_1\vtq_2}\hspace{-0.3cm}
 \tilde g(\tilde p ,\tilde q_1 ,\tilde q_2 ) \phi(\vtq_1)\phi(\vtq_2) \delta(\vtp-\vtq_1-\vtq_2) 
\ , \\ \nonumber
\Pi(\vtp)&=&
\int_{\vtq_1\vtq_2} \hspace{-0.3cm}
\tilde g(\tilde q_1 ,\tilde p ,\tilde q_2 ) \pi(\vtq_1)\phi(\vtq_2)\delta(\vtp-\vtq_1-\vtq_2)
\ , \nonumber
\end{eqnarray}
where the scale-transformed non-Gaussian variational cut-off is given by
\begin{align}
\tilde g(\tilde p ,\tilde q_1 ,\tilde q_2 )
\equiv
e^{f(\tilde{p} e^u,u)-f(\tilde q_1 e^u,u)-f(\tilde q_2 e^u,u)} \ g(\tilde p e^u,\tilde q_1 e^u,\tilde q_2 e^u) \nonumber
\ .
\end{align}
That is to say, as it occurs in the standard cMERA formulation, the variational parameters explicitly depend on the scale transformation.
Hence, the cMERA scale-dependent wavefunctional $\Psi_{NG}[\phi;u] = U_{NG}\Psi_{G}[\phi;u]$ is given by
\begin{align}
\Psi_{NG}[\phi;u]=
\Psi_G\left[\Sigma_{(-)}(\tilde p e^{u};u)\left(
	\phi(\vtp)
	-s e^{\frac d2 u}\, \Phi(\vtp) \right)\right] \nonumber
	\ ,
\end{align}
where we have assumed, for simplicity, that $\chi_0=0\,$. Regarding the solution of the Gaussian variational parameter $f(p;u)$ given in \cite{cMERA1,cMERA2}, it is straightforward to see that $\Sigma_{(\pm)}(\tilde p e^{u};u)|_{u\to 0} = 1$ and thus,  (\ref{eq:nlct_u}) reduces to  (\ref{eq:nlct}) and $\Psi_{NG}[\phi;0]= \Psi_G\left[\phi(\vp)	-s\, \Phi(\vp) \right]$. 
\\

\emph{Non-Gaussian Correlation Functions.-}
As in the Gaussian case, the non-Gaussian cMERA  {based on the $\pi\, \phi^2$} presented here is specially well suited to analyze correlation functions. These observables distinguish the ground  states of interacting theories from those of noninteracting ones: i.e., while for Gaussian states the connected correlation functions of order higher than two vanish, those of interacting systems are generally nonzero. In addition, the multiscale approach provides a procedure to gain some understanding of the non-perturbative effects taking place at different scales. 

From  (\ref{eq:nlct_u}), we write the following structure of the $n$-point correlators at scale $u$ in real space
\begin{align}
\qquad & \hspace{-20pt} G^{(n)}(\vx_1,...,\vx_n)
\equiv\
\expval{\phi_1\cdots\phi_n}_{NG}	
\nonumber
\\
=&\ 
\expval{\phi_1\cdots \phi_n}_{G}	
\nonumber\\\nonumber
&\ 
+ s\left [	
	\expval{\Phi_1\phi_2\cdots \phi_n}_{G}
	+\cdots
	+\expval{\phi_1\cdots \phi_{n-1}\Phi_n}_{G} \right] 	
\\\nonumber
&\ 	
+s^2\left[ 	
	\expval{\Phi_1\Phi_2\phi_3\cdots \phi_n}_{G}	
	+\cdots 
	+\expval{\phi_1\cdots\Phi_{n-1}\Phi_n}_{G}\right]
\\\nonumber
&\ 
\ \, \vdots
\\
&\ 	
+s^n \expval{\Phi_1\cdots\Phi_n}_{G}\ ,
\end{align}
where $\phi_i\equiv\phi(\vx_i)$ and $\Phi_j \equiv \Phi(\vx_j)$. 
The correlation functions break up into interaction-less disconnected functions and 
connected ones containing information about the interaction. The first four connected functions are
\begin{equation}
\begin{aligned}
G_c^{(1)}(\vx_1)                     
&=
s \tilde\chi_1 
\ , \\
G_c^{(2)}(\vx_1,\vx_2)           
&= 
\tilde D(12)	+ s^2 \tilde\chi_2 (12)
\ , \\
G_c^{(3)}(\vx_1,\vx_2,\vx_3) 
&= 
s[\tilde\chi_3]_{(123)}+s^3\tilde \chi_4(12,23,31)
\ ,  \\
G_c^{(4)}(\vx_1,\vx_2,\vx_3,\vx_4) &=\frac{s^2}{2} \left[\tilde \chi_5\right] +s^4\left(\left[ \tilde \chi_2\, \tilde \chi_2
\right] + \left[ \tilde \chi_6\right]\right)
\ ,
\end{aligned}
\label{connected_corr}
\end{equation}
where we use the notation $ab\equiv \vx_{ab} \equiv \vx_{a} -\vx_{b}$. 
$\tilde D(ab)\equiv D(ab;u) $ is the scale-dependent propagator
\begin{align}
\tilde D(ab) 
= 
\frac12 \int_\vp  e^{-2f(p,u)}F(p e^{-u})\, e^{i\vp \cdot \vx_{ab}}
\ .
\end{align}
The loop integrals $\tilde \chi_i(\vx;u)$, $i=1,\, \cdots\, 6$ depend both on the positions and the scale $u$ and 
 their explicit expressions and  bracketed quantities involving them can be found in \cite{supp_material}. 

Connected functions show how the non-Gaussian cMERA procedure goes beyond the Gaussian approximation and captures scale-dependent non-perturbative contributions, which are arranged in powers of the variational parameter $s$. Focusing on quantities that usually  measure the non-Gaussianity of a system, we notice that the \emph{skewness}, related with the $3$-point function, is given by {$\gamma_1^2(s;u)\equiv \frac{\left(G_c^{(3)}(123)\right)^2}{(G_c^{(2)})^3} \underset{s\to 0}{\sim}
\frac{([\tilde \chi_3]_{123})^2}{[\tilde D\tilde D\tilde D]}\, s^2 +{\cal O}(s^4),$}
{where $(G_c^{(2)})^3 \equiv G_c^{(2)}(12)G_c^{(2)}(13)G_c^{(2)}(23)$.} In the limit of large $s$ ($s\to\infty$), the skewness  achieves the limiting value $\gamma^2_{1,\infty}\sim \tilde \chi_4^2(12,23,31)/[\tilde\chi_2\tilde\chi_2\tilde\chi_2]+{\cal O}(s^{-2})$. In this sense, the quantities that usually can be measured in the experiments are the full and connected 2-point and 4-point correlation functions, as well as the point-dependent excess \emph{kurtosis} over a Gaussian model \cite{kurtosis}. For the latter, we obtain, 
\begin{align}
\gamma_2(s;u) 
\equiv&\
\frac{G_c^{(4)}(1234)}{[G_c^{(2)}G_c^{(2)}]}
\underset{s\to 0}{\sim} \frac{[\tilde \chi_5]}{2[\tilde D\tilde D]} s^2+ O(s^4)
\ ,
\end{align}
where $[G_c^{(2)}G_c^{(2)}]=G_c^{(2)}(12) G_c^{(2)}(34)+G_c^{(2)}(13) G_c^{(2)}(24)+G_c^{(2)}(14) G_c^{(2)}(23)$.
In the limit of strong non-Gaussianity, $s\to\infty$, the excess kurtosis goes to a limiting value 
{$\gamma_{2,\infty}\sim 1 + [\tilde \chi_6]/[\tilde\chi_2\tilde\chi_2]+{\cal O}(s^{-2})$.} 
\\

\emph{Equations for the variational parameters.-}
We remark that to fully {solve the non-Gaussian cMERA tensor network and} evaluate the previous expressions {for a  theory with a Hamiltonian $\cH$}, we must obtain the optimal values for the variational parameters $F(p)$, $g( p , q_1, q_2 )$ and $s$. { This is addressed by minimizing the expectation value of the energy density $\langle \cH \rangle = \langle \Psi_{G} |U_{NG}^{\dagger}\, \cH\, U_{NG} | \Psi_{G}\rangle\,$} at some length scale $u$, that in our case is the UV limit, (i.e., $u\to0$). {Here, we discuss two different theories. First we consider the $\lambda  \phi^4$ scalar theory whose Hamiltonian density reads $\cH_{\phi^4} = \cH_{\rm kin} + \frac{1}{2}\ m^2\, \phi(x)^2 + \frac{\lambda}{4!}\phi(x)^4\, ,$}
{where $\cH_{\rm kin}= 1/2\left(\pi(x)^2 + \left[\nabla \phi(x)\right]^2\right)$ and $m$ and $\lambda$ are the bare mass and the bare coupling respectively. 
As any other polynomial interaction is a straightforward extension of this work, we also discuss a non-power-like potential, such as the sine-Gordon model whose Hamiltonian is given by $\cH_{\bf sG} = \cH_{\rm kin} - \frac{\alpha}{\beta^2}\left[\cos \beta \phi(x)-1\right]\, , $}
{in which $\beta$ is a dimensionless parameter, while $\alpha$ can be
regarded as the square of the bare mass in the case of
vanishing $\beta$. The energy expectation value of the $\lambda \phi^4$ theory is }
{
\begin{align}
\label{eq:energy_functional_phi4}
\expval{\cH_{\phi^4}}
=\expval{\cH_{\rm kin}} +\frac12 m^2(s^2 \chi_2+\phi_c^2) \\
+\frac{\lambda}{4!}\Big[ 3I^2+6s^2(I\chi_2+\chi_5) +3s^4(\chi_2^2+ \chi_6)\nonumber \\
+4\phi_c(3s \chi_3+s^3 \chi_4)+6\phi_c^2(I+s^2 \chi_2)+\phi_c^4 \nonumber
\Big]  \, , 
\end{align}
}
{where $\phi_c=\chi_0+s\chi_{1}$, $\expval{\cH_{\rm kin}}=\frac{1}{4}\int_{\vp}\left[F(p)^{-1} + p^2F(p)\right] + s^2 \chi_7$  
and $I = 1/2 \int_\vp F(p)$.} The notation $\chi_{i}$ means that the loop integrals are evaluated at the 
same spatial point $\vx$, i.e., $\chi_{i}\equiv \tilde \chi_{i}(\vx_{ab}=0;u=0)$.

The equations for the optimal values of the variational parameters  $s$, $F(p)$ and $g(p,q_1,q_2)$ are
obtained, {for a fixed $\phi_c$,} by deriving $\expval{\cH_{\phi^4}}$ w.r.t. them and then equating to zero \cite{supp_material}. {This yields a set of non-linear coupled equations that must be self-consistently and numerically solved. However, our aim here is to provide expressions that explicitly show the relation between the variational parameters and the coupling constants of the models under consideration. To proceed, we note that $\expval{\cH_{\phi^4}}$ and their related optimization equations greatly simplify for $\phi_c \sim 0$ where the kernel $F(p)$ reduces to $F(p)=1/\sqrt{p^2 + \mu^2} + \mathcal{O}(\phi_c^2)$, with $\mu$ a variational parameter \cite{ibanez, supp_material}. In that case, $\mu^2 = m^2 + (\lambda/2)\, I_0(\mu^2)$, where $I_0(\mu^2)
= \frac12\int_\vp (p^2+\mu^2)^{-\frac12}$. Further, we note that only the product $s\, g$ is meaningful and thus, fixing $s$ to be $s = -4\, \lambda\, \phi_c$ is a way to conveniently normalize $g$\footnote{This relation does not impose any restriction on $\lambda$ which, in particular, can be finite.}. Finally, denoting $f(\vp,\vq) \equiv g(|\vp+\vq|,p,q)$,    the optimal cut-off function is the solution of}
\beqa
f(\vp,\vq)\hspace{-0.1cm}& =&\hspace{-0.1cm} G(\vp, \vq)
\nonumber \\ 
& \times & 
\left(1-4\lambda\int_{\vk}\left[f(\vp,\vk) + f(\vq,\vk)\right]F(k)\right)\, ,
\label{eq:SD_g}
\eeqa
{with $G(\vp, \vq)$ a combination of kernels given in \cite{supp_material}}. The term proportional to $s\, \chi_3$ in (\ref{eq:energy_functional_phi4}) is the major contribution to the improvement of the energy value compared to the Gaussian estimate \cite{polley89, ritschel90}. 
Indeed, the optimal $\chi_3$ (given in terms of the solution of (\ref{eq:SD_g}) is seen to contain 
an infinite series of diagrammatic contributions to the two-point function that are complementary to 
the ``cactus''-diagrams resummation \cite{ritschel94}. This highlights to what extent, the trial wavefunctionals of the 
non-Gaussian cMERA, may produce approximations that go far beyond the Gaussian approximation. 
Remarkably, the NLCT procedure in $d=1$ includes more physics but no further infinities than those 
posed by the cactus-diagrams. However, the renormalization of the non-Gaussian variational calculations in $d > 1$ is shown to be much more involved and the contributions generated by the NLCT need infinite rescalings of the bare parameters \cite{ritschel90}. 

{Regarding the sine-Gordon model, when computing $\langle \cH_{\bf sG}\rangle$, the term $\langle \cos \beta \phi \rangle$ poses a challenge to the NLCT method as this interaction term is non-polynomial. 
In \cite{supp_material} it is shown that when the momemtum support of the $p$-modes in $g(p,q_1,q_2)$ is sufficiently small in comparison with the support of the $q$-modes, one may write $\langle \cos \beta \phi \rangle=\exp\left(-\beta^2/2\, I_0(\mu^2)\right)\cos \beta \varphi_c$ with $\varphi_c = s \chi_1$ and $\mu$ a variational mass parameter. In this limit, the optimization procedure can be applied to this model.}

\emph{Discussion.-}
In this work, a general method for building non-Gaussian generalizations of the cMERA has been presented. The method uses a class of non-linear canonical transformations which are then applied to a Gaussian wavefunctional. 
 We have shown how to obtain non-perturbative effects on the correlation functions far beyond the Gaussian approximation in 
two scalar field theories. {We expect this can be useful in addressing recent experimental data on higher order correlation functions in many body systems \cite{kurtosis, schied_nature}}. Furthermore, our method shows how the cMERA formalism could provide a systematic UV regularization scheme for generic interacting QFTs. In this sense, our approach can be generalized to fermionic and gauge field theories. In particular, we propose the following fermionic transformation acting on a spinor $\psi(\vk)$: 
\beqa
\cB &=& \int_{\vp\vq_1\cdots\vq_l} \hspace{-0.5cm}g^{\alpha \beta_1\cdots \beta_l}(\vp,\vq_1,\cdots,\vq_l) \pi_\alpha(\vp)\\ \nonumber
&& \times\, \psi_{\beta_1}(\vq_1)\cdots \psi_{\beta_l}(\vq_l) \delta(\vp+\vq_1+\cdots+\vq_l)
\ ,
\eeqa
where Greek indices denote spinor components, $g$ is a variational (non-)Grassmannian function and $\pi_\alpha(\vp)\equiv \delta/\delta\psi_\alpha(-\vp)$ is the conjugate momentum. Despite this transformation also truncates, a model-dependent analysis, which is beyond the scope of this paper, would impose additional restrictions on the indices $\beta_i$. We expect this transformation to be useful in addressing relevant physical phenomena in strongly coupled theories including chiral field theories. 

Regarding dynamical settings such as quantum quenches, the method promises to be useful as for the moment, all studies with the Gaussian cMERA, assume that the time-evolved state after the quench remains Gaussian along the evolution. Finally, it is worth to explore what geometrical interpretation can be found for the non-Gaussian cMERA ansatz presented in this work.
\\

\emph{Acknowledgements.-} 
JJFM acknowledges A. Bhattacharyya and T. Takayanagi for useful discussions and comments.
JMV thanks J.Cotler and M.Muller for many fruitful discussions. 
The work of JJFM is supported by Universidad de Murcia. 
JMV is supported by Ministerio de Economia y Competitividad FIS2015-69512-R and Programa de Excelencia de la Fundacion Seneca 19882/GERM/15. 
JJFM and ETL  acknowledge the financial support of Spanish Ministerio de Economia y Competitividad and CARM Fundaci\'on Seneca under grants FIS2015-28521 and 21257/PI/19.
\\

\onecolumngrid
\newpage

\section*{--- \ Supplemental Material \ ---
}

\subsection{I. $\tilde \chi$ Integrals}
The loop integrals $\tilde \chi_i$ {related with the nonlinear field transformation $\cB = \pi\, \phi^2$ depend on both positions and the renormalization scale $u$. Once the optimal variational parameters $F(p)$ and $g(p,q_1,q_2)$ are obtained for a concrete theory, then higher-order correlation functions can be computed through them.} Their explicit expressions  are 
\begin{align}
\tilde \chi_1
=&\
\frac12 \int_\vp  \tilde g (0,p e^{-u},p e^{-u}) \ F(p e^{-u})
\ ,
\\
\tilde \chi_2(a\, b) 
=&\
\frac12 \int_{\vp_1\vp_2} e^{i(\vp_1+\vp_2)\cdot \vx_{ab}} e^{-2f(|\vp_1+\vp_2|,u)}
\tilde g (e^{-u}|\vp_1+\vp_2|,e^{-u}p_1,e^{-u}p_2)^2 F(p_1 e^{-u})F(p_2 e^{-u})
\ ,
\\
\tilde \chi_3(ab,cd) 
=&\
\frac12 \int_{\vp_1\vp_2} e^{i(\vp_1\cdot \vx_{ab}+\vp_2\cdot \vx_{cd}} e^{-f(p_1,u)-f(p_2,u)-f(|\vp_1+\vp_2|,u)}
\nonumber\\
&
\hspace{60pt}\times 
\tilde g (e^{-u}|\vp_1+\vp_2|,e^{-u}p_1,e^{-u}p_2)F(p_1 e^{-u})F(p_2 e^{-u})
\ ,
\\
\tilde \chi_4(ab,cd,ef) 
=&\
\int_{\vp_1\vp_2\vp_3} e^{i(\vp_1\cdot \vx_{ab}+\vp_2\cdot \vx_{cd}+\vp_3\cdot \vx_{ef})} e^{-f(|\vp_2-\vp_3|,u)-f(|\vp_3-\vp_1|,u)-f(|\vp_1-\vp_2|,u)}
\nonumber\\
&
\hspace{60pt}\times 
\tilde g (e^{-u}|\vp_1-\vp_2|,e^{-u}p_1,e^{-u}p_2)\tilde g (e^{-u}|\vp_2-\vp_3|,e^{-u}p_2,e^{-u}p_3)
\nonumber\\
&
\hspace{60pt}\times 
\tilde g (e^{-u}|\vp_3-\vp_1|,e^{-u}p_3,e^{-u}p_1) F(p_1 e^{-u})F(p_2 e^{-u})F(p_3 e^{-u})
\ ,
\\
\tilde \chi_5(ab,cd,ef) 
=&\
\int_{\vp_1\vp_2\vp_3} e^{i(\vp_1\cdot \vx_{ab}+\vp_2\cdot \vx_{cd}+\vp_3\cdot \vx_{ef})} e^{-f(|\vp_1+\vp_2|,u)-f(|\vp_2+\vp_3|,u)-f(|\vp_3+\vp_1|,u)}
\nonumber\\
&
\hspace{60pt}\times 
\tilde g (e^{-u}|\vp_1+\vp_2|,e^{-u}p_1,e^{-u}p_2)\tilde g (e^{-u}|\vp_1+\vp_3|,e^{-u}p_1,e^{-u}p_3)
\nonumber\\
&
\hspace{60pt}\times 
F(p_1 e^{-u})F(p_2 e^{-u})F(p_3 e^{-u})
\ ,
\\
\tilde \chi_6(ab,cd,ef,gh) 
=&\
\int_{\vp_1\cdots\vp_4} e^{i(\vp_1\cdot \vx_{ab}+\vp_2\cdot \vx_{cd}+\vp_3\cdot \vx_{ef}+\vp_4\cdot \vx_{gh})} e^{-f(|\vp_1+\vp_2|,u)-f(|\vp_1+\vp_3|,u)-f(|\vp_2+\vp_4|,u)-f(|\vp_4+\vp_1|,u)}
\nonumber\\
&
\hspace{60pt}\times 
\tilde g (e^{-u}|\vp_1+\vp_2|,e^{-u}p_1,e^{-u}p_2) \tilde g (e^{-u}|\vp_1+\vp_3|,e^{-u}p_1,e^{-u}p_3) 
\nonumber\\
&
\hspace{60pt}\times 
\tilde g (e^{-u}|\vp_2+\vp_4|,e^{-u}p_2,e^{-u}p_4) \tilde g (e^{-u}|\vp_3+\vp_4|,e^{-u}p_3,e^{-u}p_4)
\nonumber\\
&
\hspace{60pt}\times 
F(p_1 e^{-u})F(p_2 e^{-u})F(p_3 e^{-u})F(p_4 e^{-u})
\ .
\end{align}

In terms of these integrals, we define the following quantities in brackets:
\begin{align}
[\tilde\chi_3]_{(123)}
	=&\
	\tilde\chi_3(12,13)
	+\tilde\chi_3(13,12)
	+\tilde\chi_3(13,23) \ ,
	\nonumber
	\\[5pt]
[\tilde\chi_5]
		=&\
		\tilde\chi_5(12,32,14)
		+\tilde\chi_5(12,42,13)
		+\tilde\chi_5(13,23,14)
		+\tilde\chi_5(13,43,12)
		+\tilde\chi_5(14,24,13)
		+\tilde\chi_5(14,34,12)
		\nonumber\\
		&\
		+\tilde\chi_5(23,13,24)
		+\tilde\chi_5(23,43,21)
		+\tilde\chi_5(24,14,23)
		+\tilde\chi_5(24,34,21)
		+\tilde\chi_5(34,14,32)
		+\tilde\chi_5(34,24,31)\ ,\nonumber\\[5pt]
[\tilde\chi_6] 
		=&\
		\tilde\chi_6(12,23,34,41)
		+\tilde\chi_6(13,34,42,21)
		+\tilde\chi_6(14,23,34,41)\ ,
		\nonumber\\[5pt]
[\tilde D \tilde D ]
		=&\
		\tilde D(12)\tilde D(34)
		+\tilde D(13)\tilde D(24)
		+\tilde D(14)\tilde D(23) \ ,
		\nonumber\\[5pt]
[\tilde D \tilde D \tilde D]
		=&\
		\tilde D(12)\tilde D(13)\tilde D(23)\ ,
		\nonumber\\[5pt]
[\tilde\chi_2 \tilde\chi_2] 
		=&\
		\tilde\chi_2(12)\tilde\chi_2(34)
		+\tilde\chi_2(13)\tilde\chi_2(24)
		+\tilde\chi_2(14)\tilde\chi_2(23)\ ,\nonumber\\[5pt]
[\tilde \chi_2 \tilde\chi_2 \tilde \chi_2]
		=&\
		\tilde \chi_2(12)\tilde \chi_2(13)\tilde \chi_2(23)
		\ .
\end{align}

\subsection*{II. Optimized variational parameters}

{In this section we show some details of the optimization procedure of the non-Gaussian cMERA. As mentioned in the main text, we proceed in the UV scale.} 

\subsubsection*{The $\lambda \phi^4 theory$}
{We recall the energy functional expectation value of the $\lambda \phi^4$ theory} 
\begin{align}
\label{sm-eq:energy_functional_phi4}
\expval{\cH_{\phi^4}}
=\expval{\cH_{\rm kin}} +\frac12 m^2(s^2 \chi_2+\phi_c^2) 
+\frac{\lambda}{4!}\Big[ 3I^2+6s^2(I\chi_2+\chi_5) +3s^4(\chi_2^2+ \chi_6)
+4\phi_c(3s \chi_3+s^3 \chi_4)+6\phi_c^2(I+s^2 \chi_2)+\phi_c^4 \nonumber
\Big]  \, , 
\end{align}
{where $\phi_c=\chi_0+s\chi_{1}$ and}
\begin{align}
\expval{\cH_{\rm kin}}=\frac{1}{4}\int_{\vp}\left[F(p)^{-1} + p^2F(p)\right] + s^2 \chi_7\, .
\end{align}
{The loop integral $\chi_7$ is given by}
\beqa
\chi_7&=&
\frac14 \int_{\vp\vq} \left[	g(|\vp+\vq|,p,q)^2 |\vp+\vq|^2 F(p)F(q)+ g(p,|\vp+\vq|,q)^2 F(q) F(p)^{-1}
	\right]
\ ,
\eeqa
\\
{and the notation $\chi_{i}$ without a tilde means that the loop integrals mentioned in the previous section are evaluated at the same spatial point $\vx$, i.e., $\chi_{i}\equiv \tilde \chi_{i}(\vx_{ab}=0;u=0)$.}
\\

{The optimal values for the kernel $F(p)$ and $f(\vp,\vq)$, where, for convenience, we define $f(\vp,\vq)\equiv g(|\vp+\vq|,p,q)$, are found by setting $\delta\expval{\cH_{\phi^4}}/\delta f(\vp,\vq)$ and $\delta\expval{\cH_{\phi^4}}/\delta F(k)^{-1}$ equal to zero. Despite this can be done in full generality, $\phi_c$ has to be fixed, in order for the trial wavefunctions to be consistent with the Rayleigh-Ritz method \cite{sm-Stevenson:1985zy,sm-Sher:1988mj}. This yields a set of nonlinear coupled equations that must be solved numerically and self-consistently. These are given by (they can also be found in \cite{sm-ibanez}),}
\beqa
\frac{\delta \expval{\cH_{\phi^4}}}{\delta F(k)^{-1}} &=& \frac{\delta \expval{\cH}_{G}}{\delta F(k)^{-1}} - \frac12  F(k)^2\Bigg[s^2 \int_\vp (\vp + \vk)^2 f^2(\vp,\vk) F(p) + s^2 \int_\vp f^2(\vp,\vk)/F(|\vp + \vk|)  + s^2 \Omega^2 \int_\vp f^2(\vp,\vk) F(p) \nonumber \\
&+&  2\lambda \Bigg[ 4 \phi_c s \int_\vp f(\vp,\vk) F(p) + s^2 \chi_2 + 2 s^2 \int_{\vp,\, \vq} [2 f(\vp, \vq) f(\vq,\vk) + f(\vp, \vk) f(\vk,\, \vq)] F(p) F(q) \nonumber \\
&+& 12 \phi_c\, s^3 \int_{\vp, \, \vq} f(\vp, \vq) f(\vq,\vk) f(\vq, -\vk) F(p) F(q) + 12 s^4 \int_{\vp,\, \vq,\, \vr} f(\vp, \vq) f(\vq,\vr) f(\vq, \vk)f(\vr, \vk)F(p) F(q) F(r) \nonumber \\
&+& 6 s^4 \chi_2 \int_\vp f^2(\vp,\vk) F(p)
\Bigg]
\Bigg]\, ,
\nonumber \\
\\
\frac{\delta \expval{\cH_{\phi^4}}}{\delta f(\vp,\vq)} &=& \frac12 s^2  f(\vp,\vq) F(p) F(q)\Bigg[(\vp + \vq)^2 + \frac{1}{F(|\vp + \vq|)}\left(\frac{1}{F(p)} + \frac{1}{F(q)}\right) + \Omega^2 \Bigg] \nonumber \\ 
&+& 2 \lambda F(p) F(q)\Bigg[s\, \phi_c  
+ s^2 \int_{\vr}[f(\vp,\vr) + f(\vq, \vr)] F(r) + 6 s^3 \phi_c \int_\vr f(\vr,\vp)f(\vq, -\vr)F(r) + 3 s^4 \chi_2 f(\vp, \vq)\nonumber \\
& +& 6 s^4 \int_{\vr,\, \vs} f(\vr,\vs) f(\vr,\vp) f(\vs,\vq)F(r) F(s)\Bigg]\, ,
\eeqa
where, for later convenience, the equation for the kernel $F(k)$ has been arranged in terms of $ \expval{\cH}_{G}$, 
\begin{align}
\expval{\cH}_G
=
\frac14 \int_\vp \left(
	F(p)
	+\frac{\vp^2}{F(p)}
	\right)
+\frac12m^2 (I_0+\chi_0^2)
+\lambda\left(
	\chi_0^4
	+6I_0\chi_0^2
	+3I_0^2
	\right)
\ ,
\end{align}
the expectation value yielded by a pure Gaussian ansatz, $\Omega^2 = m^2 +\frac{\lambda}{2}(I + \phi_c^2 )$ and $I = 1/2\int_\vp\, F(p)$, $I_0 = 1/2\int_\vp\, F(p)^{-1}$.
\\

{From the above equations we note that the product $s\, f$ is really meaningful so first, we choose to fix $s$ to  $s = -4\lambda\, \phi_c$ as a way to conveniently normalize $f(\vp,\vq)$. With this, here we will focus on the solution of these equations when $\phi_c=0$. At $\phi_c = 0$, the optimization equations above greatly simplify. In Eq. (10), only the Gaussian term survives so the kernel  reduce to the Gaussian solution}
\begin{align}
F(p)=\frac{1}{\sqrt{p^2 + \mu^2}}\, ,\quad \mu^2 = m^2 + \frac{\lambda}{2}\, I_0(\mu^2)\, ,
\end{align}
{with $I_0(\mu^2) = \frac{1}{2}\int_\vp\, (p^2 + \mu^2)^{-1/2}$. With this, the cut-off function is the solution of 
\begin{align}
f(\vp,\vq)=G(\vp,\vq) \left(1-4\lambda\int_\vk\, \left[f(\vp,\vk) + f(\vq,\vr) \right]\ F(k)\right)\, ,
\end{align}
{where}
\begin{align}
G(\vp,\vq)^{-1}=\left((\vp + \vq)^2 + \mu^2  + \frac{1}{F(|\vp + \vq|)}\, \left[ \frac{1}{F(p)} + \frac{1}{F(q)}\right]\right)\, .
\end{align}

\medskip

Hence, we conclude that, upon calculating the particular loop integral appearing in $\expval{\cH}$, we have shown evidence on the validity of this method to be applied to any interacting theory with a polynomial potential. In the next section, we will go a step further and analyse a non-polynomial case.

\subsubsection*{The sine-Gordon model}

In this section we illustrate the validity of our method when applied to the sine-Gordon model. Having a non-polynomial potential, we obtain some particular non-perturbative results, as a full treatment relays beyond the scope of this letter.

{The expectation value of the energy density in the sine-Gordon model is given by}
\begin{align}
\expval{\cH_{\bf sG}}=\expval{\cH_{\rm kin}}-\frac{\alpha}{\beta^2}\left[\expval{\cos \beta \phi}-1\right]\, ,
\end{align}
{which, as commented in the main text, poses a challenge to the method of non linear canonical transformations due to the non power-like potential term $\cos \beta \phi$. Here, we show a limit for which a closed expression for $\expval{\cos \beta \phi}$ can be obtained, and thus, it is possible to apply an optimization procedure similar to the one exposed above for the $\lambda \phi^4$ theory. To this end, let us first consider the shifted Gaussian wavefunctional given by}
\beq
\Psi[\phi]_{SG} = \exp\left(-\frac{1}{2}\, \int_{\vp} \left(\phi(\vp) -\chi_0 \right) F^{-1}(p)\left(\phi(-\vp) -\chi_0  \right) \right)\, ,
\eeq
{where $F(p) = (p^2 + \mu^2)^{-1/2}$, with $\mu$ a variational mass parameter. We note that for arbitrary local even powers of $\phi$, the expectation values in $\Psi[\phi]_{SG}$ are}
\begin{align}
\expval{\phi^{2l}}_{SG}= \sum_{k=0}^{l}\binom{2l}{2k}\, \frac{2 k!}{2^k k!}\, I_0^{k}\, \chi_0^{2l - 2k}\, ,
\end{align}
{where $I_0\equiv I_0(\mu^2)= \frac12 \int_\vp F(p)  = \frac12 \int_\vp (p^2 + \mu^2)^{-1/2}$. With this, through a formal expansion of the $\cos \beta \phi$ it is easy to show that \cite{sm-ingermanson86}
}

\begin{align}
\expval{\cos \beta \phi}_{SG}= \exp \left(-\frac{\beta^2}{2}\, I_0\right)\, \cos \beta \chi_0 \, .
\end{align} 

{The relevant point for us is that, in some limit, the shifted Gaussian turns out to be a special case of the $\pi \phi^2$-transformed Gaussian. To show this, we note that after the nonlinear transformation  of a Gaussian wavefunctional $\Psi_G[\phi]$, $\expval{\phi}= s \chi_1 = \varphi_c$. Now, let us define {\bf P} and {\bf Q} to be the domain support for the transformed $p$-modes of the field and the support for the shifting $q$-modes respectively (see Eqs. (11) and (12) of the main text). Here, we consider a special limit of {\bf P} and {\bf Q}. {\bf P} is to be a sphere with volume ${\rm V}_{\bf P}$ and center at the origin. {\bf Q} is a spherical shell which surrounds {\bf P} with volume ${\rm V}_{\bf Q}$. Now we take the limit of small $\varepsilon \equiv  {\rm V}_{\bf P}/{\rm V}_{\bf Q}$. As a result, in this limit only the loop integral $\chi_1$ becomes independent of $\varepsilon$ while for
the remaining $\chi$'s we can easily obtain upper bounds which depend on $\varepsilon$. As an example we consider the integral}
\begin{align}
\chi_2 
=&\
\frac12 \int_{\vp \vq} 
g (|\vp+\vq|,p,q)^2 F(p)F(q)
\ .
\end{align} 

{The $q$-integration only contributes inside a sphere with volume ${\rm V}_{\bf P}$ and center at $- p$, with $- p \in {\bf Q}$. Thus, the $p$-integration is carried out on ${\bf Q}$ and a rough upper bound for $\chi_2$ is given by}
\begin{align}
\chi_2 <\frac12\,  {\rm V}_{\bf P}\, {\rm V}_{\bf Q}\, F^{2}_{\rm max} = \frac{\varepsilon}{2}\,   \left({\rm V}_{\bf Q}\, F_{\rm max}\right)^2\, ,
\end{align} 
{with $F_{\rm max}$ the maximum of the function $F(p)$ which is assumed to be bounded from above. For the $\chi_1$ integral we estimate}
\begin{align}
\chi_1 <\frac12\,   {\rm V}_{\bf Q}\, F_{\rm mean}\, ,
\end{align} 
{where $F_{\rm mean}$ is an intermediate value of $F(p)$. Recalling that $\expval{\phi^2} = I_0 + s^2(\chi_1^2 + \chi_2)$, from the previous estimations we write}
\begin{align}
\expval{\phi^2}= 
I_0 + s^2 \chi_1^2\left(1 + \frac{\chi_2}{\chi_1^2}\right)
< I_0 + s^2 \chi_1^2\left(1 + 2 \varepsilon\left[\frac{F_{\rm max}}{F_{\rm mean}}\right]^2\right)
 \underset{\varepsilon\to 0}{\sim} I_0 + s^2 \chi_1^2 
 = I_0 + \varphi_c^2
 \, .
\end{align} 

{The same is valid for other loop integrals entering the expectation values of higher powers of $\phi$. Thus, one
obtains in the limit $\varepsilon \to 0$:
}

\begin{align}
\expval{\phi^{2l}}_{\varepsilon}= \sum_{k=0}^{l}\binom{2l}{2k}\, \frac{2 k!}{2^k k!}\, I_0^{k}\, \varphi_c^{2l - 2k}\, ,
\end{align}

{where subscript $\varepsilon$ refers to the $\pi \phi^2$ non-Gaussian wavefunctional in the limit that has been commented above. That result is consistent with writing}

\begin{align}
\expval{\cos \beta \phi}_{\varepsilon}= \exp \left(-\frac{\beta^2}{2}\, I_0\right)\, \cos \beta\, \varphi_c \, .
\end{align} 
 
{This closed form expression for the expectation value of the cosine interaction allows us to apply an optimization procedure for the variational parameters of the sine-Gordon model similar to the one proposed for the $\lambda \phi^4$ theory.}

Hence, we have shown evidence of the validity of this method even when non-polynomial interactions are considered.

\subsection*{III. Scale-dependent correlators} 

\begin{figure}[t]
\centering
\includegraphics*[width=.451\textwidth]{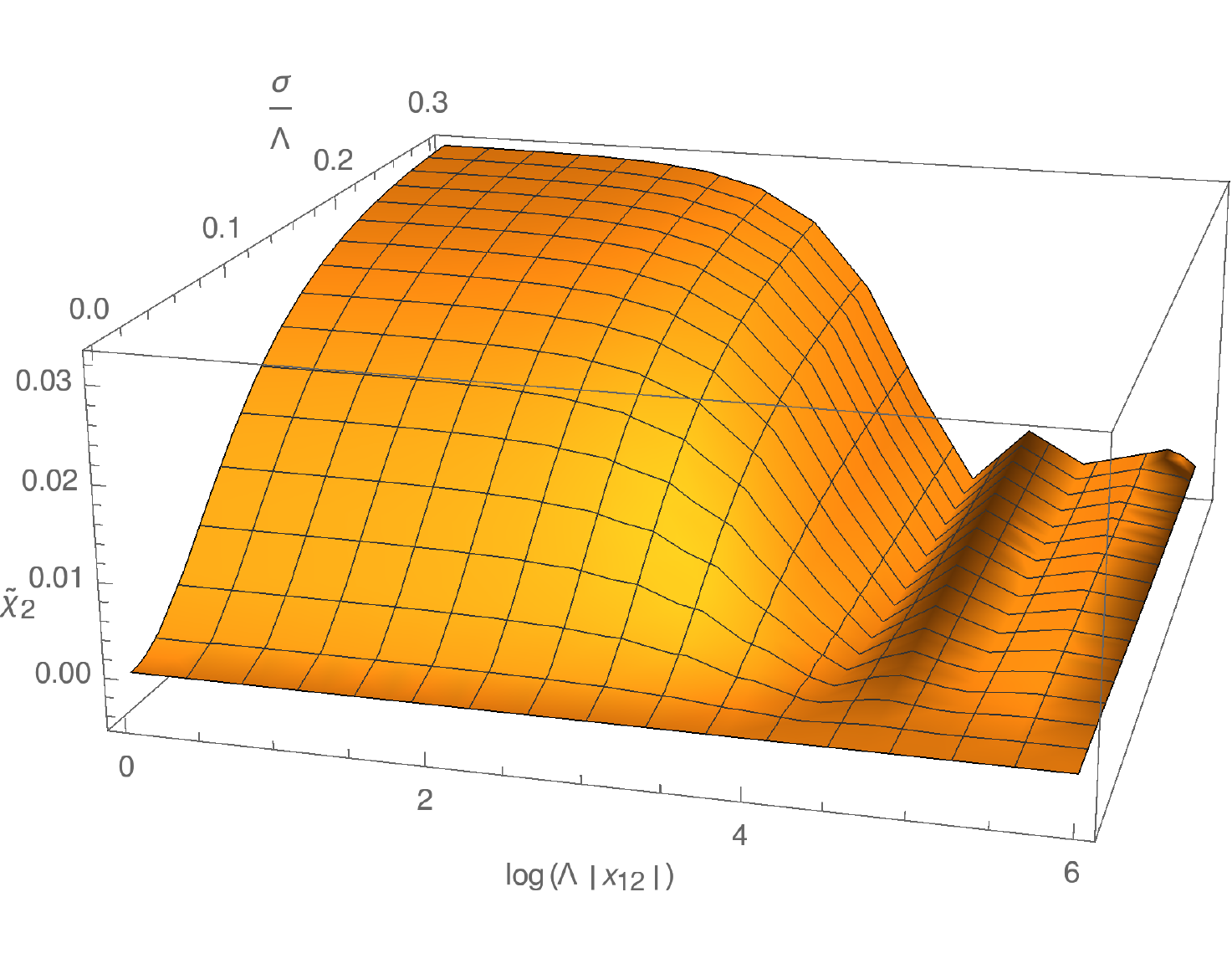}
\caption{
\textit{
The reduced 2-point correlator, $\tilde\chi_2(12)=s^{-2}\left (G_c^{(2)}(12)-\tilde D(12)\right)$, as a function 
of the scale $\sigma/\Lambda$ and the distance $|\vx_{12}|$ ($\Lambda=100$, $\mu=10$, $C_1/C_2=0.06$, the details of the optimization can be found in \cite{sm-polley89, sm-ritschel90}).}
}
\label{fig:2-point}
\end{figure}

The  correlators are modulated by the scale transformations. 
As an example, let us consider the 2-point function $G_c^{(2)}(12)$, which is given by
\begin{equation}
G_c^{(2)}(\vx_1,\vx_2)           
= 
\tilde D(12)	+ s^2 \tilde\chi_2 (12)
\ . 
\end{equation}

In Fig. \ref{fig:2-point} we show the term  $\tilde\chi_2(12)=s^{-2}\left (G_c^{(2)}(12)-\tilde D(12)\right)$ {for the $\lambda \phi^4$} as a function of the position $|\vx_{12}|$ at a given scale $u= \log \frac{\sigma}{\Lambda}$.  As expected, the non-Gaussian contributions  showed in the figure, vanish when $\sigma\to 0$.

The integral $\tilde\chi_2$ depends on the function $g(p,q,r)$ and the kernel $F(p)$. {As commented in the previous section, 
 we take the form of the latter as $F(p)=(p^2 + \mu^2)^{-1/2}$. In addition, we use an \emph{ansatz} for $g(p,q,r)$ in terms of the cMERA momentum cut-off function $\Gamma(x)$ and two variational parameters $C_{1,2}$ \cite{sm-polley89, sm-ritschel90}. Thus, we choose} 
\begin{align}
g(p,q,r)=\Gamma((p/C_1)^2)\, \left[\Gamma((C_1/q)^2)-\Gamma((C_2/q)^2)\right] \, \left[\Gamma((C_1/r)^2)-\Gamma((C_2/r)^2)\right]\, ,
\end{align} 
where  $C_{1,2}$ {can be understood as} variationally optimized coupling-dependent momentum cut-offs, with $|C_i|\le\Lambda$. With this choice, the optimal function $g(p,q,r)$  must be found self-consistently by determining the cut-offs $C_{1,2}$ which are coupling-dependent. The same applies to the remaining variational parameter $F$. 
That is to say, from a cMERA point of view, the equation above strongly suggests that $g(p,q,r)$ might be understood as a variational coupling-dependent momentum cut-off function. Upon minimization, this function potentially exhibits the non-trivial interaction effects of the theory, which turn out to be essential in the case in which the Gaussian quasi-particle picture is no longer valid.



\begin{thebibliography}{99}
\bibitem{Vidal1} G. Vidal, 
\textit{Physical Review Letters} 99, (2007) 220405,
[0512165].

\bibitem{cMERA1}
J.Haegeman, T. J. Osborne, H.Verschelde, F. Verstraete. 
\textit{Physical Review Letters} 110 (2013): 100402.

\bibitem{cMERA2}
M. Nozaki,  S. Ryu, T.Takayanagi. 
\textit{Journal of High Energy Physics} 10 (2012): 1-40.

\bibitem{Vidal2} A.Franco-Rubio, G.Vidal, 
\textit{Journal of High Energy Physics} 12 (2017):129.

\bibitem{HolCMERA1} M. Miyaji, T. Numasawa, N. Shiba, T. Takayanagi and K. Watanabe, 
\textit{Physical Review Letters} 115,(2015) 171602.

\bibitem{HolCMERA2} M. Miyaji, T. Takayanagi and K. Watanabe, 
\textit{Phys. Rev.} D 95, 066004 (2017).
1609.04645.

\bibitem{HolCMERA3} A. Mollabashi, M. Nozaki, S. Ryu and T. Takayanagi, 
\textit{Journal of High Energy Physics} 03(2014), 98, [1311.6095].

\bibitem{HolCMERA4} P. Caputa, N. Kundu, M. Miyaji, T. Takayanagi and K. Watanabe, 
\textit{Phys.Rev.Lett.} 119 (2017) no.7, 071602.

\bibitem{JMV1} J. Molina-Vilaplana, 
\textit{Journal of High Energy Physics} 03(2015). 

\bibitem{JMV2} J. Molina-Vilaplana, 
\textit{Physics Letters B}, 755 (2016) 421-425.

\bibitem{Cotler1} J. S. Cotler, J. Molina-Vilaplana and M. T. Mueller, 
[1612.02427].

\bibitem{Cotler2} J. Cotler, M. Reza Mohammadi Mozaffar, A. Mollabashi, A. Naseh, 
[1806.02835]

\bibitem{cluster2}  C.S. Hsue, H. K\"ummel, P. Ueberholz, 
\textit{Phys. Rev.} D32 (1985) 1435.

\bibitem{cluster3} U.Kaulfuss, M. Altenbokum, 
\textit{ Phys. Rev.} D35 (1987) 609

\bibitem{cluster4}  M. Funke, U.Kaulfuss, H. K\"ummel, 
\textit{Phys. Rev.} D35 (1987) 621

\bibitem{cluster5}  D. Horn, M. Weinstein, 
\textit{Phys. Rev.} D30 (1984) 1256

\bibitem{polley89} L. Polley and U. Ritschel, 
\textit{Phys. Lett.} B 221,44 (1989).

\bibitem{ritschel90} U. Ritschel, 
\textit{Zeitschrift für Physik} C 47(3):457-467 (1990).

\bibitem{ritschel91} U. Ritschel, 
\textit{Zeitschrift für Physik} C 51:469-475 (1991).

\bibitem{ritschel94} U. Ritschel, 
\textit{Zeitschrift für Physik} C 63:345-350 (1994).

\bibitem{ibanez} R. Iba\~nez-Meier, A. Mattingly, U. Ritschel, and P. M. Stevenson, 
\textit{Phys.Rev.}  D45, (1992) 15, 
 
\bibitem{stevenson} Stevenson, P. M. 
\textit{Physical Review} D 32.6 (1985): 1389.

\bibitem{Cotler3} J. Cotler, M. Reza Mohammadi Mozaffar, A. Mollabashi, A. Naseh, 
[1806.02831]

\bibitem{supp_material} See Supplemental Material for further details.

\bibitem{kurtosis}  I. Kukuljan, S. Sotiriadis and G. Takacs,
\textit{Phys. Rev. Lett. } 121, 110402 (2018)
  
\bibitem{schied_nature}  T. Schweigler, V. Kasper, S. Erne, I. Mazets, B. Rauer, F. Cataldini, T. Langen, T. Gasenzer, J. Berges ,  J. Schmiedmayer
\textit{Nature} 545, 323-326 (2017)  

\end{thebibliography}

\begin{thebibliography}{99}


\bibitem{sm-Stevenson:1985zy}
P.~M.~Stevenson,
\textit{Phys.\ Rev.\ D} {\bf 32} (1985) 1389.
doi:10.1103/PhysRevD.32.1389


\bibitem{sm-Sher:1988mj}
M.~Sher,
\textit{Phys.\ Rept.\ }  {\bf 179} (1989) 273.
  

\bibitem{sm-ibanez} R. Iba\~nez-Meier, A. Mattingly, U. Ritschel, and P. M. Stevenson, 
\textit{Phys.Rev.}  D45, (1992) 15.


\bibitem{sm-ingermanson86} R. Ingermanson 
\textit{Nuclear Physics} {\bf B266} 3-4, (1986), 620-632.


\bibitem{sm-polley89} L. Polley and U. Ritschel, 
\textit{Phys. Lett.} B 221,44 (1989).


\bibitem{sm-ritschel90} U. Ritschel, 
\textit{Zeitschrift für Physik} C 47(3):457-467 (1990).


\end{thebibliography}
\end{document}